\documentclass[12pt]{amsart}
\usepackage{comment}

\usepackage[english]{babel}
\usepackage[T1]{fontenc}

\usepackage{a4wide}

\usepackage[utf8]{inputenc}
\usepackage{amsmath,amssymb,mathrsfs}
\usepackage{amscd}
\usepackage{tikz}
\numberwithin{equation}{section}
\newtheorem{theo}{Theorem}[section]

\newtheorem{prop}{Proposition}[section]
\newtheorem{lem}{Lemma}[section]
\newtheorem{déf}{Definition}[section]
\theoremstyle{remark} 

\numberwithin{figure}{section}

\newenvironment{prof}
	{\textit{\textbf{Proof.}}}
	{\hfill $\square$\vskip 8pt}

\title[Counting function of magnetic eigenvalues]{Counting function of magnetic eigenvalues for non-definite sign perturbations}
\author{Diomba \textsc{Sambou}}
\address{Departamento de Matem\'aticas, Facultad de Matem\'aticas, 
Pontificia Universidad Cat\'olica de Chile, Vicu\~na Mackenna 
4860, Santiago de Chile}

\email{disambou@mat.uc.cl}

\thanks{This research is partially supported by the Chilean 
Program \textit{N\'ucleo Milenio de F\'isica Matem\'atica
RC$120002$}. I am grateful to J. F. Bony for 
suggesting me this study and the exploitation of the 
reduction \eqref{eq4,2}. 
}


\begin{document}

\begin{abstract}
We consider the perturbed operator $H(b,V) := H(b,0) 
+ V$, where $H(b,0)$ is the $3$d Hamiltonian of Pauli 
with non-constant magnetic field, and $V$ is 
\textit{a non-definite sign electric potential} 
decaying exponentially with respect to the variable 
along the magnetic field. We prove that the only 
resonances of $H(b,V)$ near the low ground state zero 
of $H(b,0)$ are its eigenvalues and are concentrated 
in the semi axis $(-\infty,0)$. Further, we establish 
new asymptotic expansions, upper and lower bounds on
their number near zero.
\end{abstract}

\maketitle

\section{Introduction}

In this article, we consider a three-dimensional Pauli 
operator $H(b,V) = H(b,0) + V$ acting in 
$L^{2}(\mathbb{R}^{3}) := L^{2}(\mathbb{R}^{3},
\mathbb{C}^{2})$. Its describes a quantum non-relativistic 
spin-$\frac{1}{2}$ particle, subject to an electric 
potential $V$ and a non-constant
magnetic field $\textbf{B} : \mathbb{R}^{3} \rightarrow 
\mathbb{R}^{3}$ of constant direction. With no loss of 
generality, we may assume that the magnetic field has 
the form 
\begin{equation}\label{eq1,0}
\textbf{B}(x_1,x_2,x_3) = \big( 0,0,b(x_1,x_2) \big).
\end{equation} 
Throughout this paper, $b : \mathbb{R}^{2} \rightarrow 
\mathbb{R}$ will be assumed to be an admissible 
magnetic field. That is, there exists a constant 
$b_{0} > 0$ satisfying $b(x_1,x_2) = b_{0} + 
\tilde{b}(x_1,x_2)$, where $\tilde{b}$ is a function such 
that the Poisson equation 
\begin{equation}\label{eq1,1}
\Delta \tilde{\varphi} = \tilde{b}, \quad 
\small{\Delta := \partial_1^2 + \partial_2^2,}
\end{equation}
admits a solution $\tilde{\varphi} \in C^{2}(\mathbb{R}^{2})$ 
verifying $\sup_{(x_1,x_2) \in \mathbb{R}^{2}} \vert D^{\alpha} 
\tilde{\varphi}(x_1,x_2) \vert < \infty$, 
$\alpha \in \mathbb{N}^{2}$, $\vert \alpha \vert \leq 2$, 
\big(we refer for instance to \cite[section 2.1]{raik} for 
more details on admissible magnetic fields\big). 
Notice that $\tilde{b} = 0$ coincides 
with the constant magnetic field case. 
\\

\noindent
Let $\textbf{A} = (A_{1},A_{2},A_{3}) : 
\mathbb{R}^{3} \rightarrow \mathbb{R}^{3}$ be a magnetic 
potential generating the magnetic field $\textbf{B}$. That is,
\begin{equation}\label{eq1,2}
\textbf{B}(X) = \text{curl} \hspace{0.6mm} 
\textbf{A}(X), \quad X = (X_{\perp},x_3) \in \mathbb{R}^{3}, 
\quad X_{\perp} = (x_1,x_2) \in \mathbb{R}^{2}.
\end{equation}
The self-adjoint unperturbed Pauli operator $H(b,0)$ is 
defined originally on 
$C_{0}^{\infty}(\mathbb{R}^{3},\mathbb{C}^{2})$ by 
\begin{equation}\label{eq1,3}
H(b,0) := \begin{pmatrix}
   (-i\nabla - \textbf{A})^{2} - b & 0 \\
   0 & (-i\nabla - \textbf{A})^{2} + b
\end{pmatrix},
\end{equation}
and then closed in $L^{2}(\mathbb{R}^{3})$. Since $b$ is 
independent of $x_3$, then with no loss of generality, we 
may assume that $A_{j}$, $j = 1$, $2$, are independent of 
$x_3$ and $A_{3} = 0$. Set $\varphi_{0}(X_{\perp}) := 
b_{0} \vert X_{\perp} \vert^{2}/4$ and 
$\varphi := \varphi_{0} + \tilde{\varphi}$, so that 
we have $\Delta \varphi = b$. Introduce the operators
\begin{equation}\label{eq1,4}
a = a(b) := -2i \textup{e}^{-\varphi} \frac{\partial}{\partial \bar{z}} \textup{e}^{\varphi} \quad \text{and} \quad a^{\ast} = a^{\ast}(b) := -2i \textup{e}^{\varphi} \frac{\partial}{\partial z} \textup{e}^{-\varphi},
\end{equation}
originally defined on $C_{0}^{\infty}(\mathbb{R}^{2},\mathbb{C})$,
where $z := x_{1} + i x_{2}$ and $\bar{z} := x_{1} - i x_{2}$. 
Define the operators
\begin{equation}\label{eq1,41}
H_{1}(b) := a^{\ast} a \quad \text{and} \quad H_{2}(b) :=  a a^{\ast}.
\end{equation}
By choosing $A_1 = -\partial_2 \varphi$ and 
$A_2 = \partial_1 \varphi$,  the operator $H(b,0)$ can 
be rewritten in $L^{2}(\mathbb{R}^{3}) = L^{2}(\mathbb{R}^{2}) 
\otimes L^{2}(\mathbb{R})$ as
\begin{equation}\label{eq1,42}
\small{H(b,0) = \begin{pmatrix}
   H_{1}(b) \otimes 1 + 1 \otimes \big( -\partial_3^2 \big) & 0 \\
   0 & H_{2}(b) \otimes 1 + 1 \otimes \big( -\partial_3^2 \big)
\end{pmatrix} =: \begin{pmatrix}
   \mathcal{H}_{1}(b) & 0 \\
   0 & \mathcal{H}_{2}(b)
\end{pmatrix}},
\end{equation}
where $- \partial_3^2$ is originally defined on 
$C_{0}^{\infty}(\mathbb{R}, \mathbb{C})$. From 
\cite[Proposition 1.1]{ra}, we know that the spectra 
$\textup{sp} \hspace*{0.5mm} (H_{j})$ of $H_{j}$, 
$j = 1$, $2$, satisfy the following properties:
\begin{equation}\label{eq1,5}
\begin{split}
& \textup{sp} \hspace*{0.5mm} (H_{1}) \subseteq 
\lbrace 0 \rbrace \cup [\zeta,+\infty) \hspace{0.1cm} 
\text{with $0$ an eigenvalue of infinite multiplicity}, \\
& \textup{sp} \hspace*{0.5mm} (H_{2}) \subseteq 
[\zeta,+\infty),
\end{split}
\end{equation}
where 
\begin{equation}\label{eq1,6}
\zeta := 2 b_{0} \textup{e}^{-2 \textup{osc} 
\hspace{0.5mm} \tilde{\varphi}} > 0,
\end{equation}
with $\textup{osc} \hspace{0.5mm} \tilde{\varphi}:= 
\sup_{X_\perp \in \mathbb{R}^{2}} \tilde{\varphi} 
(X_\perp) - \inf_{X_\perp \in \mathbb{R}^{2}} 
\tilde{\varphi} (X_\perp)$. 
Since the spectrum of the operator $- \partial_3^2$ 
coincides with $[0,+\infty)$ and is absolutely 
continuous, then \eqref{eq1,42} and \eqref{eq1,5} 
imply that this of $H(b,0)$ is equal to
$[0,+\infty)$ and is absolutely continuous 
\big(see \cite[Corollary 2.2]{raik}\big). 
\\

\noindent
\textbf{Remark.}
It is well known \big(see e.g. \cite{dim}\big) 
in the constant magnetic field case, the spectrum 
of $H_{1}$ consists of the Landau levels 
$2b_{0}\mathbb{N}$. Further, the multiplicity 
of each eigenvalue $2b_{0}q$, $q \in \mathbb{N}$,
is infinite. In particular, this implies that the 
spectrum of $H_{2}$ consists of the Landau levels 
$2b_{0}\mathbb{N}^{\ast}$. Further, $\zeta = 2 b_{0}$.
\\

\noindent 
On the domain of the operator $H(b,0)$, we introduce 
the perturbed operator 
\begin{equation}\label{eq1,14}
H(b,V) = H(b,0) + V,
\end{equation}
where we identify $V$ with the multiplication operator 
by the function $V$. 
\\

\noindent
In \cite{diom}, we investigated 
the resonances (see Definition \ref{de} below) of 
the operator $H(b,V)$ near zero. We required 
$V \equiv \lbrace V_{jk} \rbrace_{1\leq j,k \leq 2}$ 
to be a hermitian matrix-valued electric potential 
satisfying
\begin{equation}\label{eq1,15}
\vert V_{jk}(X) \vert \leq C \hspace{0.5mm} \langle 
X_{\perp} \rangle^{-m_{\perp}} \text{e}^{-2 \delta 
\langle x_3 \rangle}, \quad m_{\perp} > 0, \quad \delta > 0,
\end{equation}
where $\langle u \rangle : = \sqrt{1 + \vert u 
\vert^{2}}$, $u \in \mathbb{R}^{d}$, 
$d \geq 1$.  For $V$ of definite sign, we obtained 
in \cite[Theorem 2.2] {diom} an asymptotic 
expansion of the number of resonances near zero. 
Further, we showed that they are concentrated in 
some sector. For $V$ of \textit{non-definite sign}, 
we obtained in \cite[Theorem 2.1] {diom} an 
upper bound of the number of resonances near zero 
without their localization. 
\\

\noindent
The aim of this paper is to study the same problem by 
considering the class of anti-diagonal matrix-valued 
electric potentials
\begin{equation}\label{eq1,11}
V(X) := \begin{pmatrix}
   0 & \overline{U}(X) \\
   U(X) & 0
\end{pmatrix}, \quad X \in \mathbb{R}^{3}, \quad U(X) 
\in \mathbb{C},
\end{equation}
where the function $U$ satisfies the estimate
\begin{equation}\label{eq1,13}
\vert U(X) \vert \leq C \hspace{0.5mm} \langle 
X_{\perp} \rangle^{-m_{\perp}} \text{e}^{- 2\delta 
\langle x_3 \rangle}, \quad m_{\perp} > 0, \quad 
\delta > 0,
\end{equation}
with $C > 0$ a constant. 
\\

\noindent
\textbf{Remark.} Notice that potentials $V$ satisfying 
\eqref{eq1,11} are of \textit{non-definite sign}. 
Indeed, its eigenvalues are $\pm{\vert U(X) \vert}$. 
\\

\noindent
Novelty in this paper is 
that we prove the only resonances of $H(b,V)$ 
near zero are its eigenvalues. Further, they are 
localized in the semi axis $(-\infty,0)$. We 
give new estimates on the number of negative 
eigenvalues $H(b,V)$ near zero. In particular, 
they show that the behaviour of magnetic 
eigenvalues for unsigned perturbations is different 
from that for signed perturbations.
The crucial tool is 
that we exploit the form \eqref{eq1,11} of $V$ 
in such a way we reduce the analysis of the resonances 
of $H(b,V)$ near $z = 0$ to that of the semi-effective 
effective Hamiltonian $\mathcal{H}_{1} - \overline{U} \big( 
\mathcal{H}_{2} - z \big)^{-1} U$ (see Section 4). 
\\

\noindent
The paper is organized in the following manner. 
Our main results (Theorems \ref{theo1} and 
\ref{theo2}) are stated in Section 2. In Section 3, 
we recall auxiliary results on Toeplitz 
operators and characteristic values of meromorphic
operator-valued functions. In Section 4, we reduce 
the analysis of the resonances near zero to a 
characteristic value problem. Section 5 is devoted 
to the proofs of Theorems \ref{theo1} and \ref{theo2}.

\section{Statement of the main results}\label{s2}

In order to formulate our main results,
some notations are needed. 
For $T$ a linear compact self-adjoint operator 
in a Hilbert space, we denote 
\begin{equation}\label{eq2,1}
n_{+}(s,T) := \text{rank} \hspace{0.6mm} 
\mathbb{P}_{(s,\infty)}(T), \quad s > 0,
\end{equation}
where $\mathbb{P}_{(s,\infty)}(T)$ is the 
orthogonal projection of $T$ in the interval 
$(s,\infty)$. 
The set of negative eigenvalues of the operator 
$H(b,V)$ is denoted 
$\text{sp}_{\textup{disc}} \big( H(b,V) \big)$,
namely its discrete spectrum.
The orthogonal 
projection onto $\textup{Ker} \hspace{0.5mm} 
H_{1}(b)$ defined by \eqref{eq1,41} is denoted
$p := p(b)$. The corresponding
orthogonal projection in the constant magnetic
field case will be denoted $p_{0} := p(b_{0})$.
\\

\noindent
For a bounded operator $\mathscr{B} \in \mathscr{L} 
\big( L^{2}(\mathbb{R}^{3}) \big)$, we define on 
$L^{2}(\mathbb{R}^{2})$ the operator $W(\mathscr{B})$ 
by
\begin{equation}\label{eq2,2}
\big( W(\mathscr{B}) f \big) (X_{\perp}) := 
\frac{1}{2}\int_{\mathbb{R}}  \overline{U} 
(X_{\perp},x_3) \mathscr{B} (U f) (X_{\perp},x_3) dx_3,
\quad X_\perp \in \mathbb{R}^2. \\
\end{equation}
Clearly, if $I$ denotes the identity on 
$L^{2}(\mathbb{R}^{3})$, then $W(I)$ is the 
multiplication operator by the function 
\begin{equation}\label{eq2,2b}
X_\perp \longmapsto 
\frac{1}{2}\int_{\mathbb{R}} \vert U 
\vert^{2}(X_{\perp},x_3) dx_3.
\end{equation}
The function \eqref{eq2,2b} will be denoted $W(I)$ 
again. Let $\mathcal{H}_{2}$ be the operator defined by 
\eqref{eq1,42}. If $U$ satisfies \eqref{eq1,13}, then
\cite[Lemma 2.4]{raik} implies that the positive 
self-adjoint operators $p W(I) p$ and 
$p W \big( \mathcal{H}_{2}^{-1} \big) p$ are 
compact on $L^{2}(\mathbb{R}^{2})$. 
\\

\noindent
We are thus led to our first main result,
where the resonances are defined in Definition 
\ref{de} below.

\begin{theo}\label{theo1} 
Assume that \eqref{eq1,11} and \eqref{eq1,13} 
hold for $V$ and $U$ respectively. Then, there 
exists a discrete set 
$\mathcal{E} \subset \mathbb{R}^{\ast}$ such 
that for any $\nu \in \mathbb{R}^{\ast} 
\setminus \mathcal{E}$, the operator 
$H(b,\nu V)$ has the following properties:

$\textup{(i)}$ Localization: near zero, the 
resonances are its negative eigenvalues.

$\textup{(ii)}$ Asymptotic: suppose that 
$n_{+} \Big( r,p W \big( \mathcal{H}_{2}^{-1} \big) 
p \Big) \rightarrow + \infty$, $r \searrow 0$. Then,
there exists a sequence $(r_{\ell})_{\ell}$ tending 
to $0$ such that
\begin{equation}\label{eq2,5}
\# \hspace{0.3mm} \textup{sp}_{\textup{disc}} 
\big( H(b,\nu V) \big) \cap 
\big( -\infty,-r_{\ell}^2 \big)
 = n_{+} \Big( r_{\ell},p W \big( \mathcal{H}_{2}^{-1} 
\big) p \Big) \big( 1 + o(1) \big), \quad \ell 
\longrightarrow \infty.
\end{equation}

$\textup{(iii)}$ Upper-bound: let $I$ be the identity 
on $L^{2}(\mathbb{R}^{3})$. If $W(I) \leq 
U_{\perp}$ with $U_{\perp}$ satisfying the assumptions 
of Lemma \ref{lem3,3}, then 
\begin{equation}\label{eq2,6}
\# \hspace{0.3mm} \textup{sp}_{\textup{disc}} 
\big( H(b,\nu V) \big) \cap \big( -\infty,-r^2 \big)
\leq n_{+} \left( r,\frac{1}{\zeta}p W(I) p 
\right) \big( 1 + o(1) \big), \quad r \searrow 0.
\end{equation}
\end{theo}

\noindent
\textbf{Remarks.}
Notice that in virtue of Lemma \ref{lem3,3},
the right hand side of \eqref{eq2,6} implies 
that the number of negative eigenvalues of 
$H(b,\nu V)$ near zero is of order 
$\mathcal{O} \big( r^{-1/m_{\perp}} \big)$, 
$r \searrow 0$. This order is better than the 
order $\mathcal{O} \big( r^{-2/m_{\perp}} \big)$ 
obtained in \cite{diom} for general perturbations 
$V$ satisfying \eqref{eq1,15}. Otherwise, if 
the function $U_{\perp}$ is compactly supported,
then \eqref{eq2,6} and \cite[Lemma 3.4]{raik} 
imply that the number of negative eigenvalues of 
$H(b,\nu V)$ near zero is of order 
$\mathcal{O} \big( (\ln \vert \ln r \vert)^{-1} 
\vert \ln r \vert \big)$, $r \searrow 0$, 
which is similar to that from \cite{diom}.

\begin{figure}\label{fig 1}
\begin{center}
\tikzstyle{+grisEncadre}=[fill=gray!30]
\tikzstyle{blancEncadre}=[fill=white!100]
\tikzstyle{grisEncadre}=[fill=gray!20]
\tikzstyle{ddEncadre}=[densely dotted]
\begin{tikzpicture}[scale=1]

\draw [+grisEncadre] (0,0) -- (180:2) arc (180:360:2) -- cycle;
\draw [blancEncadre] (0,0) -- (0:2) arc (0:180:2) -- cycle;

\draw[->] (-2.5,0) -- (2.3,0);
\draw (2.3,0) node[right] {$\textup{Re} \hspace{0.6mm} k$};
\draw[->] (0,-2.5) -- (0,2.5);
\draw (0,2.5) node[right] {$\textup{Im} \hspace{0.6mm} k$};
\draw (0,0) -- (1.73,1);
\draw (1,0.5) node[above] {$r$};

\node at (0,0.9) {\tiny{$\times$}};
\node at (0,0.7) {\tiny{$\times$}};
\node at (0,0.5) {\tiny{$\times$}};
\node at (0,0.3) {\tiny{$\times$}};
\node at (0,0.2) {\tiny{$\times$}};

\draw [ddEncadre] [<-] (-0.1,0.6) -- (-3,1.9);
\node at (-3,2.01) {\tiny{$\text{Resonances}$}};

\end{tikzpicture}
\caption{\textup{\textbf{Resonances near $0$ with 
respect to the variable $k$:} For $r \ll 1$, 
the only resonances $z(k) = k^{2}$ of $H(b,0) + V$ 
near zero are its negative eigenvalues and they 
satisfy 
$k \in i]0,+\infty)$.}} 
\end{center}
\end{figure}

\vspace*{0.5cm}

\noindent
In the constant magnetic field case 
$\textbf{B} = (0,0,b_{0})$, we obtain, in 
additional, a lower bound of the number of 
negative eigenvalues near zero. 
\\

\noindent
Before to state our result, some additional 
notations are needed. If the function $U$ 
satisfies $U(X_{\perp},x_{3}) = U_{\perp}(X_{\perp}) 
\hspace{0.05cm} \mathcal{U}(x_{3})$, where 
$U_{\perp}$ and $\mathcal{U}$ are not 
necessarily real functions, we define
\begin{equation}\label{eq2.7}
K_{1} := \frac{\left\langle \left( 
-\partial_3^2 + 2b_{0} \right)^{-1} 
\mathcal{U},\mathcal{U} \right\rangle}{2},
\end{equation}
and
\begin{equation}\label{eq2.71}
n_{\ast} \left( \left( \frac{r}{K_{1}} 
\right)^{\frac{1}{2}},p_{0}U_{\perp}p_{0} 
\right) := n_{+} \left( \frac{r}{K_{1}}, 
\big( p_{0}U_{\perp}p_{0} \big)^{\ast} 
p_{0}U_{\perp}p_{0} \right).
\end{equation}

\begin{theo}[Lower bound]\label{theo2} 
Let the magnetic field $\textup{\textbf{B}}$ 
be constant. Assume that \eqref{eq1,11} and 
\eqref{eq1,13} 
hold for $V$ and $U$ respectively. Then, there 
exists a discrete set 
$\mathcal{E} \subset \mathbb{R}^{\ast}$ such 
that for any $\nu \in \mathbb{R}^{\ast} 
\setminus \mathcal{E}$, the following holds:

Suppose that $U(X_{\perp},x_3) = U_{\perp}(X_{\perp}) 
\hspace{0.05cm} \mathcal{U}(x_3)$. If we have
$$
n_{\ast} \left( \left( \frac{r}{K_{1}} 
\right)^{\frac{1}{2}},p_{0}U_{\perp}p_{0} 
\right) = \phi(r) \big( 1 + o(1) \big), \quad r 
\searrow 0,
$$
where the function $\phi(r)$ is as in Lemma 
\ref{lem2,21}, then
\begin{equation}\label{eq2.10}
\# \hspace*{0.3mm} \textup{sp}_{\textup{disc}} 
\big( H(b,\nu V) \big) \cap \big( -\infty,-r^2 \big) 
\geq n_{\ast} \left( \left( \frac{r}{K_{1}} 
\right)^{\frac{1}{2}},p_{0}U_{\perp}p_{0} \right) 
\big( 1 + o(1) \big), \quad r \searrow 0.
\end{equation} 
In particular, if $U_{\perp} \geq 0$ and 
satisfies the assumptions of Lemma \ref{lem3,3}, 
then
\begin{equation}\label{eq2100}
\# \hspace*{0.3mm} \textup{sp}_{\textup{disc}} 
\big( H(b,\nu V) \big) \cap \big( -\infty,-r^2 \big) 
\geq n_{+} \left( \left( \frac{r}{K_{1}} 
\right)^{\frac{1}{2}},p_{0}U_{\perp}p_{0} 
\right) \big( 1 + o(1) \big), \quad r \searrow 0.
\end{equation}
\end{theo}

\noindent
\textbf{Remarks.}
Notice that estimates \eqref{eq2100} and 
\eqref{eq2,6} imply, in the constant magnetic 
field case, the number of negative eigenvalues
of $\big( H(b,\nu V) \big)$ near $0$ 
is such that
\begin{equation}\label{eq2.11}
\begin{split}
& C_{m_{\perp}} K_{1}^{1/m_{\perp}} r^{-1/m_{\perp}} 
\big( 1 + o(1) \big) \leq \\
& \# \hspace*{0.3mm} \textup{sp}_{\textup{disc}} 
\big( H(b,\nu V) \big) \cap \big( -\infty,-r^2 \big) \\
& \leq C_{m_{\perp}} K_{2}^{1/m_{\perp}} r^{-1/m_{\perp}} 
\big( 1 + o(1) \big), \quad r \searrow 0,
\end{split}
\end{equation}
where $C_{m_{\perp}}$ is the constant defined in 
Lemma \ref{lem3,3}, and 
\begin{equation}
K_{2} := (4b_0)^{-1} \int_{\mathbb{R}} \vert 
\mathcal{U}(x_{3}) \vert^{2} dx_{3}.
\end{equation}
It is easy to check that $K_{1} < K_{2}$. On the 
other hand, the lower bound in \eqref{eq2.11} implies
that the negative eigenvalues of $H(b,\nu V)$ 
accumulate to zero. 
One can compare \eqref{eq2.11} 
with the results of \cite{raik} on the asymptotic 
of the counting function of the eigenvalues of 
$H(b,V)$ near zero, when  $V \equiv \lbrace V_{jk} 
\rbrace_{1\leq j,k \leq 2}$ has a fixed sign.  
Indeed, in \cite[Corollary 3.6]{raik}, the author 
shows that if the coefficients of the potential 
$V \geq 0$ satisfy
$$
\vert V_{jk} (X) \vert = \mathcal{O} \big(
\langle X \rangle^{-\nu} \big), \quad 
1 \leq j,k \leq 2,
$$
for some $\nu > 3$, then the behaviour near zero 
of the counting function of the negative eigenvalues 
of $H(b,V)$ is of order 
$$
\mathcal{O} \big( r^{-2/(\nu - 1)} \big) 
\big( 1 + o(1) \big), \quad r \searrow 0.
$$ 
In particular, this shows that 
the behaviour of eigenvalues for unsigned 
perturbations is different from that for signed 
perturbations.

\section{Auxiliary results}

\subsection{Some results on Berezin-Toeplitz operators by Raikov \cite{raik}, \cite{ra0}}

Consider $U_{\perp} \in L^{\infty}(\mathbb{R}^{2})$. 
The asymptotic eigenvalues of the 
Berezin-Toeplitz operator $pUp$ is the 
subject of the next lemma. An integrated 
density of states (IDS) for the operator 
$H_{1} = H_{1}(b)$ is defined as follows. 
For $X_{\perp} \in \mathbb{R}^{2}$, let 
$\chi_{T,X_{\perp}}$ be the characteristic 
function of the square 
$X_{\perp} + \left( -\frac{T}{2},\frac{T}{2} 
\right)^{2}$ with $T > 0$. Denote 
$\mathbb{P}_{I}(H_1)$ the spectral 
projection of $H_{1}$ in the interval 
$I \subset \mathbb{R}$. A non-increasing 
function $g : \mathbb{R} \longrightarrow 
[0,\infty)$ is called an IDS for
$H_{1}$ if it satisfies for any 
$X_{\perp} \in \mathbb{R}^{2}$ 
$$
g(t) = \lim_{T\rightarrow\infty} T^{-2} 
\hspace{0.5mm} \textup{Tr} 
\hspace{0.5mm} \big[ \chi_{T,X_{\perp}} 
\mathbb{P}_{(-\infty,t)}(H_{1}) 
\chi_{T,X_{\perp}} \big],
$$ 
for each point $t$ of continuity of 
$g$ \big(see e.g. \cite{raik}\big). If the 
magnetic field is constant, then there 
exists naturally an IDS for the operator 
$H_{1}$ given by 
$$
g(t) = \frac{b_{0}}{2\pi} 
\sum_{q=0}^{\infty} \chi_{\mathbb{R}_{+}} 
(t - 2 b_{0} q), \quad t \in \mathbb{R},
$$
 where $\chi_{\mathbb{R}_{+}}$ is the 
 characteristic function of $\mathbb{R}_{+}$.

\begin{lem}{\cite[Theorem 2.6]{ra0}}\label{lem3,3} 
Consider $U_{\perp} \in C^{1}(\mathbb{R}^{2})$ 
such that 
$$
0 \leq U_{\perp}(X_{\perp}) \leq C_{1} 
\langle X_{\perp} \rangle^{-\alpha}, \hspace{0.2cm} 
\vert \nabla U_{\perp}(X_{\perp}) \vert \leq C_{1} 
\langle X_{\perp} \rangle^{-\alpha-1}, 
\hspace{0.2cm} X_{\perp} \in \mathbb{R}^{2},
$$ 
where $\alpha > 0$ and $C_{1} > 0$. Assume that
 
$\bullet$ $U_{\perp}(X_{\perp}) = u_{0}(X_{\perp} 
/ \vert X_{\perp} \vert) \vert X_{\perp} 
\vert^{-\alpha} \big{(} 1 + o(1) \big{)}$ 
as $\vert X_{\perp} \vert \rightarrow \infty$, 
where $u_{0}$ is a 

\hspace*{0.2cm} continuous function on $\mathbb{S}^{1}$ 
which does not vanish identically,

$\bullet$ $b$ is an admissible magnetic field,

$\bullet$ there exists an IDS $g$ for the operator 
$H_{1}(b)$. 
\\
Then we have
$$n_{+} (s,pU_{\perp}p) = C_{\alpha} 
s^{-2/\alpha} \big{(} 1 + o(1) \big{)}, 
\hspace{0.2cm} s \searrow 0,$$
where
\begin{equation}\label{eq3,1}
C_{\alpha} := \frac{b_{0}}{4\pi} \int_{\mathbb{S}^{1}} 
u_{0}(t)^{2/\alpha} dt.
\end{equation}
\end{lem}

\subsection{Results on characteristic values by Bony-Bruneau-Raikov \cite{bo}}\label{s3,2}

Let $\mathscr{H}$ be a separable Hilbert space.
We denote $S_{\infty}(\mathscr{H})$ (resp. 
$GL(\mathscr{H})$) the set of compact (resp. 
invertible) linear operators acting in 
$\mathscr{H}$.
\\

\noindent
Let $D \subseteq \mathbb{C}$ be a connected open set, 
$Z \subset D$ be a discrete and closed subset,
$A : \overline{D} \backslash Z \longrightarrow 
GL(\mathscr{H})$ be a finite meromorphic operator-valued
function \big(see e.g. \cite[Definition 2.1]{bo}\big) 
and Fredholm at each point of $Z$. The index of $A$, 
with respect to a positive oriented contour 
$\gamma$, is defined by 
\begin{equation}\label{ind}
\textup{Ind}_{\gamma} \hspace{0.5mm} A := 
\frac{1}{2i\pi} \textup{Tr} \int_{\gamma} A'(z)A(z)^{-1} 
dz = \frac{1}{2i\pi} \textup{Tr} \int_{\gamma} A(z)^{-1} 
A'(z) dz.
\end{equation}
Here, the operator $A$ does not vanish on the 
integration contour $\gamma$.
Let $\mathcal{D}$ be a domain of $\mathbb{C}$ 
containing $0$. Consider a holomorphic operator-valued 
function $T : \mathcal{D} \longrightarrow S_{\infty}(\mathscr{H})$. 
For a domain $\Omega \subset \mathcal{D} \setminus 
\lbrace 0 \rbrace$, a complex number $z \in \Omega$ 
is said to be a \textit{characteristic value} of 
$z \mapsto \mathscr{T}(z) := I - \frac{T(z)}{z}$ if 
the operator $\mathscr{T}(z)$ is not invertible. The 
multiplicity of a characteristic value $z_{0}$ 
is defined by
\begin{equation}\label{eqa,11}
\textup{mult}(z_{0}) := \textup{Ind}_{\gamma} \big( I - 
\mathscr{T}(\cdot) \big),
\end{equation}
where $\gamma$ is a small contour positively oriented, 
containing $z_{0}$ as the unique point $z$ satisfying 
$\mathscr{T}(z)$ is not invertible.
\\

\noindent
Define
$$
\mathcal{Z}(\Omega) := \left\lbrace z \in \Omega 
: I - \frac{T(z)}{z} \hspace{0.8mm} \textup{is not 
invertible} \right\rbrace.
$$ 
If there exists $z_{0} \in \Omega$ such that 
$I - \frac{T(z_{0})}{z_{0}}$ is not invertible, then 
$\mathcal{Z}(\Omega)$ is a discrete set 
\big(see e.g. \cite[proposition 4.1.4]{go}\big). So 
we define
$$
\mathcal{N}(\Omega) := \# \mathcal{Z}(\Omega).
$$ 
Assume that $T(0)$ is self-adjoint. Introduce 
$\Omega \Subset \mathbb{C} \setminus \lbrace 0 
\rbrace$ and the sector
\begin{equation}\label{eq3,5}
\mathcal{C}_{\alpha}(a,b) := \lbrace x + iy \in 
\mathbb{C} : a \leq x \leq b, -\alpha x \leq y \leq 
\alpha x \rbrace,
\end{equation}
with $a > 0$ tending to $0$ and $b > 0$. Let 
$$
n(\Lambda) := \textup{Tr} \hspace{0.6mm} 
\mathbf{1}_{\Lambda}(T(0))
$$ 
be the number of eigenvalues of the operator 
$T(0)$ lying in the interval $\Lambda \subset 
\mathbb{R}^{\ast}$, and counted with their 
multiplicity. Denoted $\Pi_{0}$ the orthogonal 
projection onto $\textup{Ker} \hspace{0.6mm} T(0)$.

\begin{lem}\label{lem3,5} 
\textup{\cite[Corollary 3.4]{bo}} Let $T$ be as 
above and $I - T'(0) \Pi_{0}$ be invertible. Assume 
that $\Omega \Subset \mathbb{C} \setminus \lbrace 
0 \rbrace$ is a bounded domain with smooth 
boundary $\partial \Omega$ which is transverse 
to the real axis at each point of $\partial \Omega 
\cap \mathbb{R}$.

$\textup{(i)}$ If $\Omega \cap \mathbb{R} = \emptyset$, 
then $\mathcal{N}(s\Omega) = 0$ for $s$ small enough. 
This implies that the characteristic values $z \in 
\mathcal{Z}(\mathcal{D})$ near $0$ satisfy $\vert 
\textup{Im} \hspace{0.6mm} z \vert = o(\vert z \vert)$.

$\textup{(ii)}$ Moreover, if the operator $T(0)$ 
has a definite sign $(\pm T(0) \geq 0)$, then 
the characteristic values $z$ near $0$ satisfy 
$\pm \textup{Re} \hspace{0.6mm} z \geq 0$, 
respectively.

$\textup{(iii)}$ If $T(0)$ is of finite rank, 
then there are no characteristic values in a 
pointed neighbourhood of $0$. Moreover, if the 
operator $T(0) \mathbf{1}_{[0,+\infty)}(\pm T(0))$ 
is of finite rank, then there are no characteristic 
values in a neighbourhood of $0$ intersected 
with $\lbrace \pm \textup{Re} \hspace{0.6mm} 
z > 0 \rbrace$, respectively.
\end{lem}

\begin{lem}\label{lem3,6} 
\textup{\cite[Theorem 3.7]{bo}} Let $T$ be as 
above and $I - T'(0) \Pi_{0}$ be invertible. 
For $\alpha > 0$ fixed, let $\mathcal{C}_{\alpha}(r,1) 
\subset \mathcal{D}$ be defined as in \eqref{eq3,5}. 
Then, for all $\delta > 0$ small enough, there 
exists $s(\delta) > 0$ such that, for all 
$0 < s < s(\delta)$, we have
\begin{equation}\label{eq3,7}
\begin{split}
\mathcal{N} \big( \mathcal{C}_{\alpha}(r,1) \big) & 
= n \big( [r,1] \big) \big( 1 + \mathcal{O} \big( 
\delta \vert \ln \delta \vert^{2} \big) \big) \\
& + \mathcal{O} \big( \vert \ln \delta \vert^{2} 
\big) n \big( [r(1-\delta),r(1+\delta)] \big) + 
\mathcal{O}_{\delta}(1),
\end{split}
\end{equation}
where the $\mathcal{O}$'s are uniform with respect 
to $s$, $\delta$ but the $\mathcal{O}_{\delta}$ may 
depend on $\delta$.
\end{lem}

\begin{lem}\label{lem3,7} 
\textup{\cite[Corollary 3.9]{bo}} Let the assumptions 
of Lemma \ref{lem3,6} hold true. Assume that there 
exists $\gamma > 0$ such that $$n([r,1]) = O(r^{-\gamma}), 
\hspace{0.2cm} r \searrow 0,$$ and that $n([r,1])$ 
grows unboundedly as $r \searrow 0$. Then there 
exists a positive sequence $(r_{k})_{k}$ tending to 
$0$ such that 
\begin{equation}\label{eq3,6}
\mathcal{N}(\mathcal{C}_{\alpha}(r_{k},1)) = n([r_{k},1]) 
(1 + o(1)), \hspace{0.2cm} k \rightarrow \infty.
\end{equation}
\end{lem}

\begin{lem}\label{lem2,21} 
\cite[Corollary 3.11]{bo} Let the assumptions of Lemma 
\ref{lem3,6} hold true. Suppose that 
$$n \big( [r,1] \big) = \Phi(r) \big(1 + o(1)\big), 
\hspace{0.2cm} r \searrow 0,$$ 
with $\Phi(r) = r^{-\gamma}$, or $\Phi(r) = \vert 
\ln r \vert^{\gamma}$, or $\Phi(r) = \big( \ln \vert 
\ln r \vert \big)^{-1} \vert \ln r \vert$, for some 
$\gamma > 0$. Then
\begin{equation}\label{eq2,35}
\mathcal{N}\big(\mathcal{C}_{\alpha}(r,1)\big) = 
\Phi(r) \big(1 + o(1)\big), \hspace{0.2cm} r \searrow 0.
\end{equation}
\end{lem}

\section{Resonances}

From here to the end, we assume that $V$ 
and $U$ satisfy \eqref{eq1,11} and 
\eqref{eq1,13} respectively.

\subsection{A preliminary property}

We establish the main property 
allowing to reduce the study of the resonances of
$H(b,V)$ near $z = 0$ to that of the semi-effective 
Hamiltonian $\mathcal{H}_{1} - \overline{U} 
\big( \mathcal{H}_{2} - z \big)^{-1} U$.
\\

\noindent
Let $z \in \mathbb{C}$ be small enough. We have 
\begin{equation}\label{eq4,11}
\begin{split}
\big( H(b,V) & - z \big) 
\begin{pmatrix}
   1 & 0 \\
   -(\mathcal{H}_{2} - z)^{-1}U & (\mathcal{H}_{2} - z)^{-1}
\end{pmatrix} \\
& = \begin{pmatrix}
   \mathcal{H}_{1} - z - \overline{U} 
   (\mathcal{H}_{2} - z)^{-1} U & \overline{U} 
   (\mathcal{H}_{2} - z)^{-1} \\
   0 & 1
\end{pmatrix}.
\end{split}
\end{equation}
Therefore,
\begin{equation}\label{eq4,2}
H(b,V) - z \hspace{0.2cm} \text{is invertible} 
\hspace{0.1cm} \Leftrightarrow \hspace{0.1cm} 
\mathcal{H}_{1} - z - \overline{U} 
(\mathcal{H}_{2} - z)^{-1} U  \hspace{0.1cm} 
\text{is invertible}.
\end{equation}
Further,
\begin{equation}\label{eq4,3}
\begin{split}
& \small{\big( H(b,V) - z \big)^{-1} = \begin{pmatrix}
   1 & 0 \\
   -(\mathcal{H}_{2} - z)^{-1}U & (\mathcal{H}_{2} - z)^{-1} 
\end{pmatrix}} \\
& \small{\times \begin{pmatrix}
   \big( \mathcal{H}_{1} - z - \overline{U} (\mathcal{H}_{2} - z)^{-1} U \big)^{-1} & -\big( \mathcal{H}_{1} - z - \overline{U} (\mathcal{H}_{2} - z)^{-1} U \big)^{-1} \overline{U} (\mathcal{H}_{2} - z)^{-1} \\
   0 & 1
\end{pmatrix}}.
\end{split}
\end{equation}
Hence, for $z$ small enough, property \eqref{eq4,2} 
allows to reduce the non-invertibility of the 
operator $H(b,V) - z$ to that of
$\mathcal{H}_{1} - z - \overline{U} (\mathcal{H}_{2} - 
z)^{-1} U$.

\subsection{Reduction to a semi-effective problem}

Consider $z$ lying in the upper half-plane
$\mathbb{C}^{+}$. Make the change of variables
\begin{equation}\label{eq5,3}
z := z(k) = k^{2} \hspace{0.2cm} \textup{for} 
\hspace{0.2cm} k \in \mathbb{C}_{1/2}^{+} := 
\big\lbrace k \in \mathbb{C}^{+} : k^{2} \in 
\mathbb{C}^{+} \big\rbrace.
\end{equation} 
Introduce the punctured disk
\begin{equation}\label{eq2,3}
D(0,\epsilon)^{\ast} := \big\lbrace k \in 
\mathbb{C} : 0 < \vert k \vert < \epsilon 
\big\rbrace, \quad \epsilon < \min 
\hspace{0.5mm} \left( \delta,\sqrt{\zeta} 
\right),
\end{equation}
where the constants $\delta$ and  $\zeta$ are 
respectively defined by \eqref{eq1,13} and 
\eqref{eq1,6}. 

\begin{prop}\label{prop5,1} 
{\cite[Proposition 4.1]{diom}} 

\noindent
Let $R(z)$ denote the resolvent of the
operator $H(b,V)$. Then, the operator 
valued-function
$$
k \longmapsto \left( R \big( z(k) \big)
 : \textup{e}^{-\delta \langle x_3 
\rangle} L^{2}(\mathbb{R}^{3}) \longrightarrow 
\textup{e}^{\delta \langle x_3 \rangle} 
L^{2}(\mathbb{R}^{3}) \right),
$$ 
admits a meromorphic extension from 
$\mathbb{C}_{1/2}^{+} \cap D(0,\epsilon)^{\ast}$ 
to $D(0,\epsilon)^{\ast}$. We shall denote 
this extension $R(z)$ again.
\end{prop}

\begin{déf}\label{de}
We define the resonances of $H(b,V)$ near zero 
as the poles of the meromorphic extension $R(z)$. 
\end{déf}

\noindent
Set $\mathscr{R}(z) := \big( \mathcal{H}_{1} - z - 
\overline{U} (\mathcal{H}_{2} - z)^{-1} U  \big)^{-1}$
and $R_2(z) := (\mathcal{H}_{2} - z)^{-1}$. 
From \eqref{eq4,3} we deduce that
\begin{equation}\label{eq5,14}
\begin{split}
& \small{\textup{e}^{-\delta \langle x_3 \rangle} 
R(z) \textup{e}^{-\delta \langle x_3 \rangle}} \\
& \small{
= \begin{pmatrix}
  \textup{e}^{-\delta \langle x_3 \rangle} \mathscr{R}(z) 
  \textup{e}^{-\delta \langle x_3 \rangle} & - \textup{e}^{-\delta \langle x_3 \rangle} \mathscr{R}(z) \overline{U} R_2(z) \textup{e}^{-\delta \langle x_3 \rangle} \\
   - \textup{e}^{-\delta \langle x_3 \rangle} R_2(z) U \mathscr{R}(z) \textup{e}^{-\delta \langle x_3 \rangle} & \textup{e}^{-\delta \langle x_3 \rangle} R_2(z) U \mathscr{R}(z) \overline{U} R_2(z) \textup{e}^{-\delta \langle x_3 \rangle} + \textup{e}^{-\delta \langle x_3 \rangle} R_2(z) \textup{e}^{-\delta \langle x_3 \rangle}
\end{pmatrix}
}.
\end{split}
\end{equation}
This together with Proposition \ref{prop5,1} 
and assumption \eqref{eq1,13} show that the poles 
of $R(z)$ coincide with those of $\mathscr{R}(z)$. 
Then, near $z = 0$, the investigation of the 
resonances of $H(b,V)$ is reduced to that of 
the semi-effective Hamiltonian 
$\mathcal{H}_{1} - \overline{U} 
(\mathcal{H}_{2} - z)^{-1} U$.

\subsection{Study of the semi-effective problem}

With the help of the decomposition
\begin{equation}\label{eq5,5}
(\mathcal{H}_{2} - z)^{-1} = \mathcal{H}_{2}^{-1} 
\Big( 1 - z\mathcal{H}_{2}^{-1} \Big)^{-1} = 
\mathcal{H}_{2}^{-1} \displaystyle\sum_{k \geq 0} 
z^{k} \mathcal{H}_{2}^{-k},
\end{equation}
$z$ being sufficiently small, we obtain
\begin{equation}\label{eq5,6}
(\mathcal{H}_{2} - z)^{-1} = \mathcal{H}_{2}^{-1/2} 
\Big( \mathcal{H}_{2}^{-1/2} + \mathcal{H}_{2}^{-1/2} 
M(z) \Big),
\end{equation}
where 
\begin{equation}\label{eq5,7}
M(z) := z \displaystyle\sum_{k \geq 0} 
z^{k} \mathcal{H}_{2}^{-k - 1}.
\end{equation}
So, \eqref{eq5,6} implies that
\begin{equation}\label{eq5,8}
\overline{U} (\mathcal{H}_{2} - z)^{-1} U = 
\overline{U} \mathcal{H}_{2}^{-1/2} 
\Big( \mathcal{H}_{2}^{-1/2} U + \mathcal{H}_{2}^{-1/2} 
M(z) U \Big).
\end{equation}
Now define the operator 
\begin{equation}\label{eq5,9}
\textbf{w} := \mathcal{H}_{2}^{-1/2} U.
\end{equation}
Thus, putting together \eqref{eq5,8} and 
\eqref{eq5,9} we obtain
\begin{equation}\label{eq5,11}
\overline{U} (\mathcal{H}_{2} - z)^{-1} U = 
\textbf{w}^{\ast} \big( 1 + M(z) \big) 
\textbf{w}.
\end{equation}

\noindent
We therefore have proved the following

\begin{lem}\label{lem5,4} 
For $z$ small enough, the operator 
$\overline{U} (H_{2} - z)^{-1} U$ 
admits the representation 
\begin{equation}\label{eq5,12}
\overline{U} (\mathcal{H}_{2} - z)^{-1} U = 
\textbf{\textup{w}}^{\ast} \big( 1 + M(z) \big) 
\textbf{\textup{w}}.
\end{equation} 
Further, the operator-valued function 
$z \longmapsto M(z)$
is analytic near $z = 0$.
\end{lem}

\noindent
Let $R_{1}(z)$ denote the resolvent of the 
operator $\mathcal{H}_{1}$. Under the 
notations of Lemma \ref{lem5,4}, the 
following lemma holds:

\begin{lem}\label{lem5,5} 
For $z$ small enough, the operator valued-function
$$
D(0,\epsilon)^{\ast} \ni k \longmapsto 
\mathcal{T}_{V} \big( z(k) \big) := \Big( 1 
+ M \big( z(k) \big) \Big) \textup{\textbf{w}} 
R_{1} \big( z(k) \big) \textup{\textbf{w}}^{\ast},
$$ 
is analytic with values in 
$S_{\infty} \left( L^{2}(\mathbb{R}^{3}) \right)$.
\end{lem}

\noindent
\begin{prof}
The analyticity of $\mathcal{T}_{V} \big( z(k) \big)$ 
holds since $M \big( z(k) \big)$ and $R_{1} \big( z(k) 
\big)$ are well defined and analytic for $k \in 
D(0,\epsilon)^{\ast}$. 
\\

\noindent
The compactness of $\mathcal{T}_{V} \big( z(k) \big)$ 
follows from that of $U R_{1} \big( z(k) \big) \overline{U}$, 
using the diamagnetic inequality and 
\cite[Theorem 2.13]{sim}.
\end{prof}

\noindent
We have the following characterization of
the resonances.

\begin{prop}\label{prop5,2} 
For $k$ near zero, the following assertions 
are equivalent: 

$\textup{(i)}$ $z(k) = k^{2}$ is a resonance 
of $H(b,V)$,

$\textup{(ii)}$ $1$ is an eigenvalue of 
$\mathcal{T}_{V} \big( z(k) \big)$.
\end{prop}

\noindent
\begin{prof}
The equivalence follows directly from the identity 
\begin{equation}\label{eq5,16}
\Big( I - \big( 1 + M(z) \big) \textup{\textbf{w}} 
R_{1} (z) \textup{\textbf{w}}^{\ast} \Big) 
\Big( I + \big( 1 + M (z) \big) \textup{\textbf{w}} 
\mathscr{R}(z) \textup{\textbf{w}}^{\ast} \Big) = I,
\end{equation}
and the fact that the poles of ${R}(z)$ coincide 
with those of $\mathscr{R}(z)$.
\end{prof}

\noindent
So, the multiplicity of a resonance 
$z := z(k)$ is defined by 
\begin{equation}\label{eq5,15}
\textup{mult}(z) := \textup{Ind}_{\gamma} 
\hspace{0.5mm} \Big( I - \mathcal{T}_{V}\big( z(\cdot) 
\big) \Big),
\end{equation}
where $\gamma$ is a small positively oriented 
contour containing $k$ as the unique point 
satisfying $z(k)$ is a resonance of $H(b,V)$ 
\big(see \eqref{ind}\big). 
\\

\noindent
Using the terminology of characteristic value 
recalled in Subsection \ref{s3,2}, Proposition 
\ref{prop5,2} can be formulated as follows:

\begin{prop}\label{prop6,2} 
For $k$ near zero, the following assertions are 
equivalent.

$\textup{(i)}$ $z = z(k)$ is a resonance of 
$H(b,V)$,

$\textup{(ii)}$ $k$ is a characteristic value of 
$I - \mathcal{T}_{V} \big( z(\cdot) \big)$.
\\
Further, according to \eqref{eq5,15}, the 
multiplicity of the resonance $z(k)$ coincides 
with this of the characteristic value 
$k$.
\end{prop}

\section{Proof of the main results}

First, let us introduce some tools. For 
$p = p(b)$, set $q := I - p$. Define on 
$L^{2}(\mathbb{R}^{3})$ the projections 
$P := p \otimes 1$ and $Q := q \otimes 1$.
If $z$ lies in the resolvent set the 
operator $\mathcal{H}_{1}$, we have 
\begin{equation}\label{eq5,1}
\begin{split}
(\mathcal{H}_{1} - z)^{-1} & = (\mathcal{H}_{1} 
- z)^{-1} P + (\mathcal{H}_{1} - z)^{-1} Q \\
& = p \otimes \mathcal{R}(z) + (\mathcal{H}_{1} 
- z)^{-1} Q,
\end{split}
\end{equation}
where the resolvent $\mathcal{R}(z) := 
\big( -\partial_3^2 - z \big)^{-1}$ 
admits the integral kernel 
\begin{equation}\label{eq5,2}
\mathcal{N}_{z}(x_3 - x_3') = i 
\textup{e}^{i \sqrt{z} 
\vert x_3 - x_3' \vert} / (2 \sqrt{z}), \quad 
\textup{Im} \hspace{0.5mm} \sqrt{z} > 0.
\end{equation}

\subsection{Proof of Theorem \ref{theo1}}

\subsubsection{Preliminary results}

Firstly, we need to split the operator 
$\mathcal{T}_{V} \big( z(k) \big)$ of Lemma 
\ref{lem5,5} with the help of 
\eqref{eq5,1}. We get
\begin{equation}\label{eq6,1}
\begin{split}
\mathcal{T}_{V} \big( z(k) \big) = 
\textup{\textbf{w}} & p \otimes \mathcal{R}(k^{2}) 
\textup{\textbf{w}}^{\ast} +
M \big( z(k) \big)  
\textup{\textbf{w}} p \otimes \mathcal{R}(k^{2})  
\textup{\textbf{w}}^{\ast} \\
& + \Big( 1 + M \big( z(k) \big) \Big) 
\textup{\textbf{w}} R_{1} \big( z(k) \big) 
Q \textup{\textbf{w}}^{\ast}.
\end{split}
\end{equation}
The operators $M \big( z(k) \big)$ and $R_{1} \big( 
z(k) \big) Q$ are analytic near zero. Then, it 
is not difficult to see that the third term of 
the right hand side of \eqref{eq6,1} is holomorphic 
near zero, with values in 
$S_{\infty} \left( L^{2}(\mathbb{R}^{3}) \right)$. 
By \eqref{eq5,2}, the integral kernel of 
$N(k) := \textup{e}^{-\delta \langle x_3 \rangle} 
\mathcal{R} (k^{2}) \textup{e}^{-\delta \langle x_3 
\rangle}$ is given by 
\begin{equation}\label{eq5,4}
\textup{e}^{-\delta \langle x_3 \rangle} 
\frac{i \textup{e}^{i k \vert x_3 - x_3' \vert}}{2 k} 
\textup{e}^{-\delta \langle x_3' \rangle}.
\end{equation}
This together with \eqref{eq5,7} imply that the 
second term of the right hand side of \eqref{eq6,1}
is analytic in a vicinity of zero, with values in 
$S_{\infty} \left( L^{2}(\mathbb{R}^{3}) \right)$.
\\

\noindent 
Now let us focus on the first term 
$\textup{\textbf{w}} p \otimes \mathcal{R}(k^{2}) 
\textup{\textbf{w}}^{\ast}$. According to 
\eqref{eq5,4}, we can write
\begin{equation}\label{eq6,2}
N(k) = \frac{1}{k}a + b(k),
\end{equation}
where $a : L^{2}(\mathbb{R}) 
\longrightarrow L^{2}(\mathbb{R})$ is the rank-one 
operator defined by 
\begin{equation}\label{eq6,3}
a(u) := \frac{i}{2} \langle u,\textup{e}^{-\delta 
\langle \cdot \rangle} \rangle \textup{e}^{-\delta 
\langle x_3 \rangle},
\end{equation}
and $b(k)$ is the Hilbert-Schmidt operator 
\big(for $k \in D(0,\epsilon)^{\ast}$\big) with 
integral kernel
\begin{equation}\label{eq6,4}
\textup{e}^{-\delta \langle x_3 \rangle} i 
\frac{ \textup{e}^{ i k \vert x_3 - x_3' \vert} - 1}{2 k} 
\textup{e}^{-\delta \langle x_3' \rangle}.
\end{equation}
Thus,
\begin{equation}\label{eq6,5}
\begin{aligned}
\textup{\textbf{w}} p \otimes \mathcal{R}(k^{2}) 
\textup{\textbf{w}}^{\ast} = \frac{i}{k} \times 
\frac{1}{2} \textup{\textbf{w}} \big( p \otimes \tau \big) 
\textup{\textbf{w}}^{\ast} + \textup{\textbf{w}} 
\big( p \otimes s(k) \big) \textup{\textbf{w}}^{\ast},
\end{aligned}
\end{equation}
where $\tau$ and $s(k)$ are operators acting from 
$\textup{e}^{-\delta \langle x_3 \rangle} L^{2}(\mathbb{R})$ 
to $\textup{e}^{\delta \langle x_3 \rangle} L^{2}(\mathbb{R})$, 
with integral kernels respectively given by $1$ and
\begin{equation}\label{eq6,6}
\frac{ 1 - \textup{e}^{ i k \vert x_3 - x_3' \vert}}{2 i k}.
\end{equation}

\noindent
We therefore have proved the following

\begin{prop}\label{prop6,1} 
Let $k \in D(0,\epsilon)^{\ast}$. Then,
\begin{equation}\label{eq6,7}
\mathcal{T}_{V} \big( z(k) \big) = i 
\frac{\textup{\textbf{w}} (p \otimes \tau) 
\textup{\textbf{w}}^{\ast}}{2k} + B(k), 
\end{equation}
where
\begin{equation}\label{eq6,8}
\begin{split}
B(k) := \textup{\textbf{w}} \big( & p \otimes s(k) 
\big) \textup{\textbf{w}}^{\ast} +
M \big( z(k) \big) \textup{\textbf{w}} p \otimes 
\mathcal{R}(k^{2}) \textup{\textbf{w}}^{\ast} \\
& + \Big( 1 + M \big( z(k) \big) \Big) 
\textup{\textbf{w}} R_{1} \big( z(k) \big) Q 
\textup{\textbf{w}}^{\ast},
\end{split}
\end{equation} 
is holomorphic in 
$D(0,\epsilon) := D(0,\epsilon)^{\ast} \cup \lbrace 0 
\rbrace$, with values in 
$S_{\infty} \left( L^{2}(\mathbb{R}^{3}) \right)$.
\end{prop}

\noindent
Notice that $\textup{\textbf{w}} (p \otimes \tau) 
\textup{\textbf{w}}^{\ast}$ is a positive self-adjoint 
compact operator. Indeed, if we define $e_{\pm}$ as 
the multiplication operators by the functions 
$e_{\pm} : x_3 \longmapsto \textup{e}^{\pm \delta 
\langle x_3 \rangle}$, it is easy to check that 
\begin{equation}\label{eq6,81}
\textup{\textbf{w}} (p \otimes \tau) 
\textup{\textbf{w}}^{\ast} = \textup{\textbf{w}}e_{+} 
(p \otimes c^{\ast} c) e_{+}\textup{\textbf{w}}^{\ast} 
= \big((p \otimes c) e_{+} \textup{\textbf{w}}^{\ast} 
\big)^{\ast} \big((p \otimes c) e_{+} 
\textup{\textbf{w}}^{\ast}\big).
\end{equation} 
Here, $c : L^{2}(\mathbb{R}) \longrightarrow \mathbb{C}$ 
is defined by $c(f) := \langle f,e_{-} \rangle$, so that 
$c^{\ast} : \mathbb{C} \longrightarrow L^{2}(\mathbb{R})$ 
is given by $c^{\ast}(\lambda) = \lambda e_{-}$. Now, with 
the help of \eqref{eq6,81}, we deduce that 
\begin{equation}\label{eq6,9}
\small{n_{+} \left( r,\frac{\textup{\textbf{w}} 
(p \otimes \tau) \textup{\textbf{w}}^{\ast}}{2} \right) 
= n_{+} \left( r,\frac{(p \otimes c) e_{+} 
\textup{\textbf{w}}^{\ast} \textup{\textbf{w}}e_{+} 
(p \otimes c^{\ast})}{2} \right), \quad r > 0,}
\end{equation} 
where the quantity $n_{+}(r,\cdot)$ is defined by 
\eqref{eq2,1}. By the definition \eqref{eq5,9} of 
$\textup{\textbf{w}}$, we have $\textup{\textbf{w}}^{\ast} 
\textup{\textbf{w}} = \overline{U} \mathcal{H}_{2}^{-1} U$.
This together with $\textup{sp} (\mathcal{H}_{2}) \subseteq 
[\zeta,+\infty)$ imply that 
\begin{equation}\label{eq6,10}
\frac{(p \otimes c) e_{+}\textup{\textbf{w}}^{\ast} 
\textup{\textbf{w}}e_{+} (p \otimes c^{\ast})}{2} = 
p W \big( \mathcal{H}_{2}^{-1} \big) p \leq 
\frac{p W (I) p}{\zeta},
\end{equation} 
where for $\mathscr{B} \in \mathscr{L} \big( 
L^{2}(\mathbb{R}^{3}) \big)$, $W(\mathscr{B})$ is the 
operator defined by \eqref{eq2,2}. Then, by combining 
\eqref{eq6,9} with \eqref{eq6,10} we obtain
\begin{equation}\label{eq6,11}
\begin{split}
n_{+} \left( r,\frac{\textup{\textbf{w}} (p \otimes \tau) 
\textup{\textbf{w}}^{\ast}}{2} \right) & = n_{+} 
\Big( r,p W \big( \mathcal{H}_{2}^{-1} \big) p \Big) \\
& \leq n_{+} \left( r,\frac{p W (I) p}{\zeta} \right),
 \quad r > 0.
\end{split}
\end{equation}

\noindent
Otherwise, according to Proposition \ref{prop6,2}, 
the study of the resonances $z(k) = k^{2}$ 
of $H(b,\nu V)$ near zero, is reduced to that 
of the characteristic values of the operator
$$
I - \mathcal{T}_{\nu V} \big( z(k) \big) 
= I + \nu^{2} \frac{T(ik)}{ik}.
$$
Here, taking into account Proposition \ref{prop6,1}, 
$T(ik) := \frac{\textup{\textbf{w}} (p \otimes \tau) 
\textup{\textbf{w}}^{\ast}}{2} - ikB(k)$ so that 
$T(0) = \frac{\textup{\textbf{w}} (p \otimes \tau) 
\textup{\textbf{w}}^{\ast}}{2}$. Let $\Pi_{0}$ be 
the orthogonal projection onto $\textup{Ker} \hspace{0.6mm} 
T(0)$. Since $T'(0) \Pi_{0}$ is compact, then, there 
exists a sequence $(\nu_{n})_{n}$ such that 
$I - \nu T'(0) \Pi_{0}$ is invertible for 
any $\nu \in \mathbb{R \setminus} \lbrace 
\nu_{n}, n \in \mathbb{N} \rbrace$. Note that 
we can take $\nu_{n} = \lambda_{n}^{-1}$, 
where $\lbrace \lambda_{n}, n \in \mathbb{N} \rbrace$ 
is the set of eigenvalues of the operator $T'(0) \Pi_{0}$.

\subsubsection{Back to the proof of Theorem \ref{theo1}}

Notations are those from Subsection 3.2.
\\

$\text{(i)}$: its follows immediately from 
Lemma \ref{lem3,5} with 
$z = -ik/\nu^{2}$.
\\

$\text{(ii)}$: Theorem \ref{theo1} $\text{(i)}$ 
shows, in particular, for $\vert k \vert$ small 
enough the resonances $z(k) = k^{2}$ are concentrated 
in the sector 
$\big\lbrace k \in D(0,\epsilon)^{\ast} : 
-ik/\nu^{2} \in \mathcal{C}_{\alpha}(r,r_{0}) 
\big\rbrace$, for any $\alpha > 0$. Hence, if 
$\textup{Res} \big( H(b,\nu V) \big)$ 
denotes the set of resonances of $H(b,\nu V)$,
we have
\begin{equation}\label{eq6,13}
\begin{split}
\# \big\lbrace z(k) & = k^{2} \in \textup{Res} \big( 
{H}(b,\nu V) \big) : r < \vert k \vert \leq 
r_{0} \big\rbrace \\
& = \# \big\lbrace z(k) = k^{2} \in \textup{Res} 
\big( H(b,\nu V) \big) : -ik/\nu^{2} 
\in \mathcal{C}_{\alpha}(r,r_{0}) \big\rbrace + 
\mathcal{O}(1) \\
& = \mathcal{N} \big( \mathcal{C}_{\alpha}(r,r_{0}) 
\big) + \mathcal{O}(1), \quad r \searrow 0.
\end{split}
\end{equation}
On the other hand, we have
\begin{equation}\label{eq6,131}
n \big( [r,r_{0}] \big) = \textup{Tr} \hspace{0.6mm} 
\mathbf{1}_{[r,r_{0}]} \big( T(0) \big) = n_{+} 
\left( r,\frac{\textup{\textbf{w}} (p \otimes \tau) 
\textup{\textbf{w}}^{\ast}}{2} \right)  + \mathcal{O}(1).
\end{equation}
This together with the inequality in \eqref{eq6,11} 
imply that
$$
n \big( [r,r_{0}] \big) \leq n_{+} 
\left( r,\frac{pW(I)p}{\zeta} \right) + \mathcal{O}(1).
$$
Then, Theorem \ref{theo1} $\text{(ii)}$  follows 
from \eqref{eq6,13} together with Lemma 
\ref{lem3,7}, \eqref{eq6,131} and the equality 
in \eqref{eq6,11}.
\\

$\text{(iii)}$: if we have $W(I) \leq U_\perp$, with 
$U_\perp$ satisfying the assumptions of Lemma 
\ref{lem3,3}, then
\begin{equation}\label{eq6.15}
n_{+} \left( r,\frac{pW(I)p}{\zeta} \right) = 
C_{m_{\perp}} (\zeta r)^{-1/m_{\perp}} \big( 1 + o(1) 
\big) , \hspace{0.2cm} r \searrow 0,
\end{equation}
where $m_{\perp}$ is the constant defined by 
\eqref{eq1,13}. Similarly to the inequality in 
\eqref{eq6,11}, we can show that
\begin{equation}\label{eq6.16}
n \big( [r,r_{0}] \big) \leq \textup{Tr} \hspace{0.6mm} 
\mathbf{1}_{[r,r_{0}]} \left( \frac{pW(I)p}{\zeta} \right) 
=: \tilde{n} \big( [r,r_{0}] \big).
\end{equation}
Note that due to \eqref{eq6.15}, 
\begin{equation}\label{eq6.160}
\tilde{n} \big( [r,r_{0}] \big) = C_{m_{\perp}} 
(\zeta r)^{-1/m_{\perp}} \big( 1 + o(1) \big), 
\quad r \searrow 0.
\end{equation}
Now if $\phi (r) = r^{-\gamma}$, $\gamma > 0$, then 
$\phi \big( r(1 \pm \nu) \big) = r^{-\gamma} 
(1 \pm \nu)^{-\gamma} = \phi (r) \big( 1 + 
\mathcal{O}(\nu) \big)$. If $\tilde{n} \big( [r,1] \big) 
= \phi (r) \big( 1 + o(1) \big)$ with $\phi \big( r(1 \pm 
\delta) \big) = \phi (r) \big( 1 + o(1) +  
\mathcal{O}(\delta) \big)$, then 
\begin{equation}\label{eq6.161}
\tilde{n} \big( \big[ r(1 - \nu),r(1 + \nu) \big] \big) 
= \tilde{n} \big( [r,1] \big) \big( o(1) + \mathcal{O}(\nu) 
\big).
\end{equation}

\noindent
Then, Theorem \ref{theo1} $\text{(iii)}$  
follows from \eqref{eq6,13} together with Lemma 
\ref{lem3,6}, \eqref{eq6.16}, \eqref{eq6.160} 
and \eqref{eq6.161}.

\subsection{Proof of Theorem \ref{theo2}}

To obtain \eqref{eq2.10}, it suffices to 
prove that if the function $U$ satisfies 
$U(X_{\perp},x_3) = U_{\perp}(X_{\perp}) 
\hspace{0.05cm} \mathcal{U}(x_3)$, then the 
following operator inequality holds:
\begin{equation}\label{eq6.17}
K_{1} \big( p_{0} U_{\perp} p_{0} \big)^{\ast} 
\big( p_{0} U_{\perp} p_{0} \big) \leq p_{0} W 
\big( \mathcal{H}_{2}^{-1} \big) p_{0}.
\end{equation}
Indeed, if \eqref{eq6.17} is true, then with 
respect to the constant magnetic field, 
the quantity
$n \big( [r,r_{0}] \big) = \textup{Tr} 
\hspace{0.6mm} \mathbf{1}_{[r,r_{0}]} \left( 
\frac{\textup{\textbf{w}} (p_{0} \otimes \tau) 
\textup{\textbf{w}}^{\ast}}{2} \right) = 
\textup{Tr} \hspace{0.6mm} \mathbf{1}_{[r,r_{0}]} 
\Big( p_{0} W \big( \mathcal{H}_{2}^{-1} \big) 
p_{0} \Big)$ satisfies
\begin{equation}\label{eq6.18}
n_{\ast} \big( [r,r_{0}] \big) := \textup{Tr} 
\hspace{0.6mm} \mathbf{1}_{[r,r_{0}]} \left[ K_{1} 
\big( p_{0} U_{\perp} p_{0} \big)^{\ast} 
\big( p_{0} U_{\perp} p_{0} \big) \right] \leq 
n \big( [r,r_{0}] \big).
\end{equation}
Further, if we have
\begin{align*}
n_{\ast} \left( \left( \frac{r}{K_{1}} 
\right)^{\frac{1}{2}},p_{0}U_{\perp}p_{0} \right) 
& := n_{+} \left( \frac{r}{K_{1}}, 
\big( p_{0}U_{\perp}p_{0} \big)^{\ast} p_{0}U_{\perp}p_{0} 
\right) \\
& = \phi(r) \big( 1 +o(1) \big), \quad r \searrow 0,
\end{align*}
where the function $\phi(r)$ is as in 
Lemma \ref{lem2,21}, then 
\begin{equation}\label{eq6.181}
n_{\ast} \big( [r,r_{0}] \big) = \phi(r) 
\big( 1 +o(1) \big), \quad r \searrow 0.
\end{equation}
Thus, \eqref{eq2.10} follows by arguing as in 
the proof of Theorem \ref{theo1} (iii)
above. 
\\

\noindent
Now let us proof \eqref{eq6.17}. If the magnetic 
field is constant, then $\mathcal{H}_{2}$ satisfies 
$$ 
\mathcal{H}_{2}^{-1} \geq  \mathcal{H}_{2}^{-1} p_{0}
 = p_{0} \otimes \big( - \partial_3^2 + 2b_{0} 
 \big)^{-1}.$$
This together with the definition \eqref{eq2,2} 
of $W \big( \mathcal{H}_{2}^{-1} \big)$ imply that,
if $U(X_{\perp},x_3) = U_{\perp}(X_{\perp}) 
\hspace{0.05cm} \mathcal{U}(x_3)$, then for any 
$f \in L^{2} \big( \mathbb{R}^{2} \big)$
\begin{equation}\label{eq6.19}
\Big\langle W \big(  \mathcal{H}_{2}^{-1} \big) f,f 
\Big\rangle \geq K_{1}\big\langle 
\overline{U_{\perp}}p_{0}U_{\perp}f,f \big\rangle.
\end{equation}
This means that we have the operator inequality
$$
W \big(  \mathcal{H}_{2}^{-1} \big) \geq K_{1} 
\overline{U_{\perp}}p_{0}U_{\perp}.
$$
Thus,
$$
p_{0} W \big(  \mathcal{H}_{2}^{-1} \big) p_{0} 
\geq K_{1} \big( p_{0} U_{\perp} p_{0} \big)^{\ast} 
\big( p_{0} U_{\perp} p_{0} \big),
$$
which is exactly \eqref{eq6.17}. This concludes the proof of Theorem \ref{theo2}.



\end{document}